\documentclass[prd,twocolumn,nofootinbib]{revtex4-2}

\usepackage{graphicx}
\usepackage{color}
\usepackage{amsmath,amsfonts,amssymb}
\usepackage{physics}
\usepackage{bbold}
\usepackage{t1enc}
\usepackage{latexsym} 
\usepackage{cancel}
\usepackage{epsfig}
\usepackage{multirow}

\DeclareMathSymbol{\shortminus}{\mathbin}{AMSa}{"39}
\DeclareMathSymbol{\shm}{\mathbin}{AMSa}{"39}

\newcommand{\app}{a_{1 1}}
\newcommand{\azz}{a_{0 0}}
\newcommand{\amm}{a_{{\textstyle \scriptscriptstyle -}1 {\textstyle \scriptscriptstyle -}1}}
\newcommand{\smo}{{\textstyle \scriptscriptstyle -}1}

\begin{document}

\title{Momentum entanglement at colliders: the $H \to WW,ZZ$ case}
\author{J. A. Aguilar-Saavedra}
\affiliation{Instituto de F\'\i sica Te\'orica, IFT-UAM/CSIC, c/ Nicol\'as Cabrera 13-15, 28049 Madrid}

\begin{abstract}
We address momentum entanglement in Higgs decays to weak boson pairs, $H \to WW,ZZ$, by discretising momentum space. The momenta of the two weak bosons are entangled, as well as the degrees of freedom in momentum and spin spaces.
For $H \to ZZ$ in the four-lepton final state, we also estimate the statistical sensitivity of entanglement measures at the Large Hadron Collider and future upgrades. The discretisation method introduced here is broadly applicable, offering a framework for studies of momentum entanglement at the energy frontier.
\end{abstract}

\maketitle

\section{Introduction}

The study of quantum correlations at the energy frontier has emerged as a promising avenue to test quantum foundations at extreme regimes. Collider experiments provide a unique environment in which quantum correlations are not only present but can become experimentally accessible, offering new probes of the quantum structure of scattering amplitudes and potential windows to physics beyond the Standard Model (SM). Quantum entanglement has received a special attention, as it is one of the essential features that distinguishes quantum mechanics from classical physics~\cite{Schrodinger:1935}.
Most collider-based investigations to date have focused on spin-spin entanglement, namely for top-quark pairs~\cite{Afik:2020onf,Severi:2021cnj,Afik:2022kwm,Aoude:2022imd,Aguilar-Saavedra:2022uye,Severi:2022qjy,Dong:2023xiw,Han:2023fci,Maltoni:2024tul,Maltoni:2024csn,Cheng:2024btk,Aoude:2025ovu},  weak-boson pairs~\cite{Aguilar-Saavedra:2022wam,Ashby-Pickering:2022umy,Aguilar-Saavedra:2022mpg,Fabbrichesi:2023cev,Morales:2023gow,Aoude:2023hxv,Bernal:2023ruk,Bernal:2024xhm,Ruzi:2024cbt,Grossi:2024jae,Wu:2024ovc,Bernal:2025zqq,DelGratta:2025qyp,Ding:2025mzj,Goncalves:2025qem,Goncalves:2025xer}, lepton pairs~\cite{Altakach:2022ywa,Ehataht:2023zzt,Fabbrichesi:2024wcd,Han:2025ewp,Zhang:2025mmm}, $b$-quark pairs~\cite{Afik:2025grr}, as well as particles of different spin like $tW$ pairs~\cite{Aguilar-Saavedra:2023hss,Aguilar-Saavedra:2024fig,Aguilar-Saavedra:2024hwd}. A few works have extended this picture by also considering orbital angular momentum (o.a.m.)~\cite{Aguilar-Saavedra:2024vpd,Aguilar-Saavedra:2024whi}. Yet, entanglement with continuous degrees of freedom, such as momentum, remains unexplored.

The study of momentum entanglement faces two challenges. First, momentum is a continuous degree of freedom. In finite-dimensional systems the Peres-Horodecki criterion~\cite{Peres:1996dw, Horodecki:1997vt} provides a practical sufficient condition to establish entanglement between two subsystems, by testing the positivity of the partially-transposed density operator. Here, in order to overcome the difficulty of dealing with a continuous variable, we use a momentum discretisation. This framework is presented in section~\ref{sec:2} for a general case. Entanglement in the discretised space can easily be established, and in turn implies  entanglement in the underlying momentum space. 

The second challenge concerns the determination of the density operator when momentum degrees of freedom are involved. At colliders, particle detection effectively projects onto momentum eigenstates, preventing direct access to off-diagonal elements with different momenta. This contrasts to spin degrees of freedom, whose interference terms can be determined from angular distributions, i.e.\ through quantum tomography, by assuming that the SM correctly describes particle decays. For the example $H \to WW,ZZ$ analysed here, assuming the quantum field theory (QFT) coherent description of the decay allows to express the density operator in terms of experimentally measurable quantities. This strategy, described in section~\ref{sec:3}, builds upon the method employed to study entanglement between spins and o.a.m.~\cite{Aguilar-Saavedra:2024whi}.
The determination of the full density operator involving (discretised) momenta and spins allows to test the entanglement between any two subsystems. In section~\ref{sec:4} we present calculations within the SM, and study the dependence on the bin size.

It is also of interest to address the experimental observability of momentum entanglement. For $H \to ZZ$, we estimate in section~\ref{sec:5} the statistical sensitivity at the Large Hadron Collider (LHC), using Run 2$+$3 data, and at its high-luminosity upgrade (HL-LHC). Under the mild assumption of CP conservation in the $H \to ZZ$ decay, the sensitivity for various entanglement measures  of the coherent QFT state exceeds the $3\sigma-5\sigma$ significance, depending on the pair of subsystems considered.

\section{Aggregated representation of density operators}
\label{sec:2}

Let us consider a density operator $\rho$ acting on a product Hilbert space $\mathcal{H}_A \otimes \mathcal{H}_B$. Let us further assume $\mathcal{H}_B$ can be written as direct sum over a finite number of subspaces,
\begin{equation}
\mathcal{H}_B = \mathcal{H}_{B_1} \oplus \mathcal{H}_{B_2} \oplus \dots \oplus  \mathcal{H}_{B_n}  \,.
\end{equation}
We consider an orthornormal basis $\{ |\phi_i \rangle \}$ for $\mathcal{H}_B$, such that $\{ |\phi_i \rangle \}_{i \in I_\alpha}$ are bases for $\mathcal{H}_{B_\alpha}$, $\alpha = 1,\dots,n$. For finite-dimensional $\mathcal{H}_B$, $I_\alpha$ are subsets of indices; for a continuous variable, $I_\alpha$ are slices in parameter space.\footnote{For example, if we parameterise three-momentum in spherical coordinates $\vec p = q (\sin \theta \cos \varphi, \allowbreak \sin \theta \sin \varphi, \allowbreak \cos \theta)$, each $I_\alpha$ can correspond to a range in the variables $q,\theta,\varphi$.} We can define a map $R: \mathcal{L}(\mathcal{H}_A \otimes \mathcal{H}_B) \to \mathcal{L}(\mathcal{H}_A \otimes \mathbb{C}^n)$ by
\begin{equation}
R(\rho) = \sum_{\alpha,\beta = 1}^{n} \sum_{\substack{m \in I_\alpha \\ n \in I_\beta}} (\mathbb{1}_A \otimes \langle \phi_m | ) \,\rho\, (\mathbb{1}_A \otimes | \phi_n \rangle ) \otimes |\alpha \rangle \langle \beta | \,.
\end{equation} 
(For continuous variables the second sum is replaced by an integral.) The action of $R$ is to perform partial traces in the subspaces $\mathcal{H}_{B_1}$, \dots, $\mathcal{H}_{B_n}$, so as to aggregate the degrees of freedom in each subspace $\mathcal{H}_{B_\alpha}$ into a single one. It is a generalisation of the partial trace over the full space $\mathcal{H}_B$. Let us label as $\{ |\psi_i\rangle \}$ a basis for $\mathcal{H}_A$. For a density operator
\begin{equation}
\rho = \sum_{ijkl} \rho_{ij}^{kl} |\psi_i \rangle \langle \psi_j | \otimes |\phi_k \rangle \langle \phi_l | \,,
\label{ec:rho}
\end{equation}
 the operator $\bar \rho = R(\rho)$ with aggregated degrees of freedom is
\begin{equation}
\bar \rho = \sum_{i,j} \sum_{\alpha, \beta = 1}^{n}  {\bar \rho}_{ij}^{\alpha \beta} |\psi_i \rangle \langle \psi_j | \otimes |\alpha \rangle \langle \beta | \,,
\label{ec:rhoR}
\end{equation}
with
\begin{equation}
{\bar \rho}_{ij}^{\alpha \beta} = \sum_{\substack{m \in I_\alpha \\ n \in I_\beta}} \rho_{ij}^{mn} \,.
\label{ec:rhoRel}
\end{equation}
For continuous variables the sums over $k,l$ in (\ref{ec:rho}) and $m,n$ in (\ref{ec:rhoRel}) are replaced by integrals, and the action of $R$ is to effectively discretise $\mathcal{H}_B$ into a n-dimensional space $\bar{\mathcal{H}}_B \cong \mathbb{C}^n$. It is straightforward to verify that if $\rho$ is Hermitian and positive semidefinite, so is $R(\rho)$, so it represents a valid density operator in $\mathcal{L}(\mathcal{H}_A \otimes \mathbb{C}^n)$.

Let us now assume $\rho$ represents a separable state, that is, it can be written as a convex sum
\begin{equation}
\rho = \sum_s p_s \rho_A^s \otimes \rho_B^s \,,
\end{equation}
with $p_s \geq 0$. Then, $\bar \rho$ is also separable. In order to explicitly show this, we focus on finite-dimensional spaces for simplicity in the notation. If we write $\rho$ as
\begin{equation}
\rho = \sum_s p_s \left( \sum_{i,j} \rho_{ij}^s |\psi_i \rangle \langle \psi_j | \right) \otimes 
\left( \sum_{k,l} \rho_{kl}^s |\phi_k \rangle \langle \phi_l | \right) \,,
\end{equation}
then
${\bar \rho}_{ij}^{\alpha \beta} = \sum_s p_s \rho_{ij}^s \rho_{\alpha \beta}^s$, with $\rho_{\alpha \beta}^s = \sum_{\substack{m \in I_\alpha \\ n \in I_\beta}} \rho_{mn}^s $, and
\begin{equation}
\bar \rho = \sum_s p_s \left( \sum_{i,j} p_{ij}^s |\psi_i \rangle \langle \psi_j | \right) \otimes
\left( \sum_{\alpha, \beta = 1}^{n}  \rho_{\alpha \beta}^s |\alpha \rangle \langle \beta | \right)
\label{ec:rhoRsep}
\end{equation}
is also separable. Consequently, provided we establish that $\bar \rho$  represents an entangled state by using the sufficient Peres-Horodecki criterion, this implies that $\rho$ also corresponds to an entangled state.

The marginalised operators $\rho_A$ (obtained by taking the partial trace of $\rho$ over $\mathcal{H}_B$) and ${\bar \rho}_A$ (partial trace of $\bar \rho$ over $\mathbb{C}^n$) are not equal. The reason is precisely that the mapping of $\mathcal{H}_B$ into $\mathbb{C}^n$ combines different degrees of freedom into single bins $\alpha = 1,\dots,n$, and the subsequent trace involves interference terms that are otherwise absent when directly tracing $\rho$ over $\mathcal{H}_B$. In our case of interest of momentum discretisation, the difference between these two operators can be exploited to assess the accuracy of the discretisation.

\section{Spin-momentum density operator in $H \to WW,ZZ$.}
\label{sec:3}

Let us begin by deriving the form of the momentum-spin density operator for $H \to V_1V_2$, $V=W,Z$.
In the standard $S$-matrix description of QFT, the $H \to V_1V_2$ decay produces a coherent superposition of the kinematically allowed momentum and spin configurations, with coefficients given by the corresponding decay amplitudes. For fixed values of the weak boson masses $m_{V_1}$, $m_{V_2}$, the decay produces a pure state
\begin{align}
|\Psi \rangle = & i (2\pi)^4 \int \frac{d^3 \vec p_1}{(2\pi)^3 2 E_1} \int \frac{d^3 \vec p_2}{(2\pi)^3 2 E_2} \notag \\
& \sum_{m_1 m_2} A_{m_1 m_2}(\vec p_1,\vec p_2) \delta^4(P - p_1 - p_2) |\vec p_1 \vec p_2 m_1 m_2 \rangle \,,
\label{ec:QFTdec}
\end{align}
where $(E_i, \vec p_i)$ are the four-momenta of $V_i$ and $P$ the Higgs boson momentum. As usual, we work in the Higgs rest frame, $P = (m_H,\vec{0} )$.  $A_{m_1 m_2}$ are the `canonical' decay amplitudes, written using a fixed $\hat z$ spin quantisation axis, with $S_z$ eigenvalues $m_1$, $m_2$ for the two weak bosons. Using polar coordinates
\begin{equation}
 \vec p_i = p_i (\sin \theta_i \cos \varphi_i, \sin \theta_i \sin \varphi_i, \cos \theta_i) \,,
 \label{ec:p}
 \end{equation}
and bearing in mind that
\begin{align}
\delta^3(\vec p - \vec{p'}) & = \frac{1}{p^2 \sin \theta} \delta(p-p') \delta(\theta - \theta') \delta(\phi - \phi') \notag \\
& = \frac{1}{p^2} \delta(p-p') \delta(\cos \theta - \cos \theta') \delta(\varphi - \varphi') \,,
\end{align}
the integral over $p_2$ can be performed with the radial $\delta$ function, leaving
\begin{align}
|\Psi \rangle & = i (2\pi)^4  \int \frac{p_1^2 d p_1}{(2\pi)^6 4 E_1 E_2} \int d\Omega_1 d\Omega_2 \notag \\
& \sum_{m_1 m_2} A_{m_1 m_2}(p_1,\Omega_1) \delta(m_H - E_1 - E_2) \notag \\
& \times \delta(\Omega_1 + \Omega_2)  |\vec p_1 \vec p_2 m_1 m_2 \rangle \,,
\end{align}
with the usual definition $d\Omega_i = d\! \cos \theta_i d\varphi_i$. We have also defined the shorthand
\begin{equation}
\delta(\Omega_1 + \Omega_2) = \delta(\cos \theta_1 + \cos \theta_2) \delta(\varphi_1 - \varphi_2 - \pi) \,,
\end{equation}
 which forces the two momenta to be back-to-back. The integral on $p_1$ can be done with the first $\delta$ function using
\begin{equation}
\delta(f(x)) = \sum_n \frac{\delta(x-x_n)}{|f'(x_n)|} \,,
\end{equation}
where $x_n$ are the zeroes of $f(x)$. In our case, the $\delta$ function sets $p_1$ to the value `$q$' such that $E_1(q) + E_2(q) = m_H$. Note that this value of $q$ depends on the weak boson masses $m_{V_1}$, $m_{V_2}$, which we have considered fixed.
Then, the state is
\begin{align}
|\Psi \rangle & = \frac{i q}{(2\pi)^2 4 m_H} \int d\Omega_1 d\Omega_2 \sum_{m_1 m_2} A_{m_1 m_2}(\Omega_1) \notag \\
& \times \delta(\Omega_1 + \Omega_2) |\vec p_1 \vec p_2 m_1 m_2 \rangle \,.
\end{align}
We refrain from performing the integral on $\Omega_2$, but instead keep these degrees of freedom in order to write the spin-momentum density operator. For this pure state it reads, up to normalisation,
\begin{align}
\tilde \rho & = \int d\Omega_1 d\Omega_2 d\Omega_1' d\Omega_2' 
\sum_{\substack{m_1 m_2 \\ m_1' m_2'}} A_{m_1 m_2}(\Omega_1) A_{m_1' m_2'}(\Omega_1')^* \notag \\
& \times 
\delta(\Omega_1 + \Omega_2)  \delta(\Omega_1' + \Omega_2') 
|\vec p_1 \vec p_2 m_1 m_2 \rangle \langle \vec{p_1}' \vec{p_2}' m_1' m_2' | \,.
\label{ec:rhoslice}
\end{align}
From now on, we use a tilde to denote unnormalised density operators, that is, not necessarily with unit trace.

The form of the canonical amplitudes $A_{m_1 m_2}$ entering the density operator (\ref{ec:rhoslice}) can be established on general grounds.
The helicity amplitudes for a scalar decay into two spin-1 particles, as is the case of $H \to V_1V_2$, $V=W,Z$, can be written down by using the formalism of Jacob and Wick~\cite{Jacob:1959at}, relying only on angular momentum conservation.\footnote{For an off-shell vector boson as in $H \to VV$, the propagator includes a scalar degree of freedom. Here we will be interested in decays into light charged leptons $\ell =e,\mu$ and neutrinos, and when coupled to massless external fermions the scalar component vanishes~\cite{Berge:2015jra}; therefore, we can safely consider the off-shell bosons as spin-1 particles. Identical-particle effects in $H \to ZZ$ are avoided by selecting $Z$ decays into different lepton flavours.}
For the two-body decay of a particle with spin $J$ and third spin component $M$, helicity amplitudes have the form
\begin{equation}
A_{\lambda_1 \lambda_2} (\theta,\varphi) = a_{\lambda_1 \lambda_2} D_{M \lambda}^{J\,*} (\varphi,\theta,0) \,,
\label{ec:aJW}
\end{equation}
where $\lambda_{1,2}$ are the helicities of the decay products `$1$' and `$2$', and $\lambda = \lambda_1 - \lambda_2$; $a_{\lambda_1 \lambda_2}$ are independent of the angles but depend on $q$, and $D^j_{m'm}(\alpha,\beta,\gamma)$ are the Wigner functions
\begin{equation}
D^j_{mm'} \equiv \langle j m' | e^{-i \alpha J_3} e^{-i \alpha J_2} e^{-i \gamma J_3} | j m\! \rangle \,.
\end{equation}
In our case $J = M = 0$ for a scalar decay, and there is no angular dependence in the helicity amplitudes.
The canonical amplitudes  $A_{m_1 m_2}$ entering the density operator (\ref{ec:rhoslice}) can then be obtained by a rotation of the polarisation vectors~\cite{Aguilar-Saavedra:2024whi}. They read
\begin{align}
A_{11}(\Omega) & = - \sqrt{\frac{2\pi}{15}} (\app + 2 \azz + \amm) \, Y_2^{-2}(\Omega) \,, \notag \\
A_{10}(\Omega) & = \sqrt{\frac{\pi}{15}} (\app + 2 \azz + \amm)  \, Y_2^{-1}(\Omega) \notag \\
& + \sqrt{\frac{\pi}{3}} (\app - \amm) \, Y_1^{-1}(\Omega) \,, \notag \\
A_{1-1}(\Omega) & = - \sqrt{\frac{2\pi}{45}} (\app + 2 \azz + \amm)  \, Y_2^{0}(\Omega)  \notag \\
& - \sqrt{\frac{\pi}{3}} (\app - \amm)  \, Y_1^{0}(\Omega) \notag \\
& - \sqrt{\frac{4\pi}{9}} (\app - \azz + \amm)  \, Y_0^{0}(\Omega) \,, \notag \\
A_{01}(\Omega) & = \sqrt{\frac{\pi}{15}} (\app + 2 \azz + \amm) \, Y_2^{-1}(\Omega)  \notag \\
& - \sqrt{\frac{\pi}{3}} (\app - \amm) \, Y_1^{-1}(\Omega) \,, \notag \\
A_{00}(\Omega) & = - \sqrt{\frac{4\pi}{45}} (\app + 2 \azz + \amm) \, Y_2^{0}(\Omega)  \notag \\
& + \sqrt{\frac{4\pi}{9}} (\app - \azz + \amm) \, Y_0^{0}(\Omega) \,, \notag \\
A_{0-1}(\Omega) & = \sqrt{\frac{\pi}{15}} (\app + 2 \azz + \amm) \, Y_2^{1}(\Omega)  \notag \\
& + \sqrt{\frac{\pi}{3}} (\app - \amm) \, Y_1^{1}(\Omega) \,, \notag \\
A_{-11}(\Omega) & = - \sqrt{\frac{2\pi}{45}} (\app + 2 \azz + \amm) \, Y_2^{0}(\Omega)  \notag \\
& + \sqrt{\frac{\pi}{3}} (\app - \amm) \, Y_1^{0}(\Omega) \notag \\
& - \sqrt{\frac{4\pi}{9}} (\app - \azz + \amm) \, Y_0^{0}(\Omega) \,, \notag \\
A_{-10}(\Omega) & = \sqrt{\frac{\pi}{15}} (\app + 2 \azz + \amm) \, Y_2^{1}(\Omega)  \notag \\
& -  \sqrt{\frac{\pi}{3}} (\app - \amm) \, Y_1^{1}(\Omega) \,, \notag \\
A_{-1-1}(\Omega) & = - \sqrt{\frac{2\pi}{15}} (\app + 2 \azz + \amm) \, Y_2^{2}(\Omega) \,,
\end{align}
with $Y_l^m(\Omega)$ the spherical harmonics. In this way, the density operator in the product Hilbert space $\mathcal{H}_{P_1} \otimes \mathcal{H}_{P_2} \otimes \mathcal{H}_{S_1} \otimes \mathcal{H}_{S_2}$ is determined.

The dependence of the helicity amplitudes $\app$, $\azz$, $\amm$ on $q$ is mild, and a single bin is sufficient for this variable. For the angular variables we divide the $\theta$ and $\varphi/2$ ranges into $\kappa= 2$, $3$, or $4$ bins, chosen so that the two-dimensional bins cover equal solid angles. For $\varphi$ the bin size is $\pi/\kappa$, while for $\theta$ the bins are
$[\arccos (1-2(i-1)/\kappa), \arccos (1-2i/\kappa) ]$, $i=1,\dots,\kappa$. The discretised momentum spaces have dimensions 8, 18 and 32, respectively. As an example, for the coarser selection $\kappa = 2$ the two-dimensional bins are
\begin{align}
& I_1 : \theta\in [0,\pi/2] \,,\; \phi \in [0,\pi/2] \,, \notag \\
& I_2 : \theta\in [0,\pi/2] \,,\; \phi \in [\pi/2,\pi] \,,  \notag \\
& I_3 : \theta\in [0,\pi/2] \,,\; \phi \in [\pi,3\pi/2] \,, \notag \\
& I_4 : \theta\in [0,\pi/2] \,,\; \phi \in [3\pi/2,2\pi] \,, \notag \\
& I_5 : \theta\in [\pi/2,\pi] \,,\; \phi \in [0,\pi/2] \,, \notag \\
& I_6 :  \theta\in [\pi/2,\pi] \,,\; \phi \in [\pi/2,\pi] \,, \notag \\
& I_7 :  \theta\in [\pi/2,\pi] \,,\; \phi \in [\pi,3\pi/2] \,,  \notag \\
& I_8 :  \theta\in [\pi/2,\pi] \,,\;  \phi \in [3\pi/2,2\pi] \,.
\end{align}
The matrix elements of the unnormalised discretised operator are obtained by integrating 
the products of amplitudes in (\ref{ec:rhoslice}) over the appropriate intervals, 
\begin{align}
{\tilde{\bar \rho}}_{m_1 m_2 m_1^\prime m_2^\prime}^{\alpha \beta \alpha' \beta'} & = \int_{I_\alpha} d\Omega_1 \int_{I_{\alpha'}} d\Omega_1^\prime A_{m_1 m_2} (\Omega_1) A_{m_1^\prime m_2^\prime} (\Omega_1')^* \notag \\
& \times \delta_{\alpha \beta} \delta_{\alpha'\beta'}  \,.
\label{ec:MEun}
\end{align}
The discretised indices $\alpha$, $\alpha'$ correspond to the integration regions for $\Omega_1$ and $\Omega_1^\prime$, respectively. The Dirac deltas in (\ref{ec:rhoslice}) become Kronecker deltas in (\ref{ec:MEun}).\footnote{In the discretised space $\bar{\mathcal{H}}_{P_2}$ we have reshuffled bin indices so as to keep a more compact expression for the Kronecker deltas.} The normalised operator $\bar \rho$ is obtained from ${\tilde{\bar \rho}}$ dividing by its trace.

In order to obtain the SM prediction for $\bar{\rho}$ covering the full decay phase space we use Monte Carlo calculations of $gg \to H \to ZZ \to e^+ e^- \mu^+ \mu^-$ and $gg \to H \to W^+W^- \to \ell^+ \nu \ell^- \nu$ with {\scshape Madgraph}~\cite{Alwall:2014hca}  at LO, using $7 \times 10^6$ and $10^7$ events, respectively. 
In $H \to ZZ$, we label the boson with largest invariant mass as $V_1$, and in $H \to W^+ W^-$ we select $V_1 = W^+$. In any case, the predictions are symmetric under interchange $1 \leftrightarrow 2$.
The value of $q$, namely, the modulus of the Higgs rest-frame three-momenta, mainly depends on the invariant mass of the off-shell boson $m_{V^*}$.  For the numerical computation of ${\bar \rho}$ we divide the $m_{V^*}$ range in 2 GeV intervals and, within each bin `$k$' of $m_{V^*}$, the values of $\app$, $\azz$ and $\amm$ are extracted from Monte Carlo pseudo-data (see the appendix for details) using parton-level information.\footnote{This slicing is more than sufficient; for 5 GeV intervals the numerical results are the same, with differences in the range $10^{-4}-10^{-3}$.}
The density operator ${\bar \rho}^{(k)}$ for that bin is calculated using (\ref{ec:MEun}), with subsequent normalisation, and the theoretical prediction for ${\bar \rho}$ in the full $m_{V^*}$ range is obtained by summing the operators ${\bar \rho}^{(k)}$ in the different $m_{V^*}$ bins, with the appropriate weights. This procedure effectively produces a discretised operator  ${\bar \rho}$ with a single bin of $q$ which, as aforementioned, is sufficient for our purposes.

\section{Entanglement in the SM}
\label{sec:4}

The entanglement between one of the subsystems and the rest can be tested by taking the partial transpose of ${\bar \rho}$ over its corresponding space. 
For a bipartite system $AB$ described by a density operator $\rho$, the entanglement can be characterised by
the Peres-Horodecki criterion. Let us label as $\rho^{T_B}$ the operator that results by taking the transpose over the $B$ space in $\rho$. (The result is the same when taking the partial transpose on subsystem $A$.)  Because the positivity of $\rho^{T_B}$ is a necessary condition for separability~\cite{Peres:1996dw, Horodecki:1997vt}, a non-positive $\rho^{T_B}$ is a sufficient condition for entanglement. Furthermore, the amount of entanglement can be quantified by the negativity of $\rho^{T_B}$~\cite{Plenio:2007zz},
\begin{equation}
N(\rho) = \frac{\| \rho^{T_B} \| - 1}{2} \,,
\label{ec:Nrho}
\end{equation}
where $\|X\| = \tr \sqrt{XX^\dagger} = \sum_i \sqrt{\lambda_i}$, where $\lambda_i$ are the (positive) eigenvalues of the matrix $XX^\dagger$. Equivalently, $N(\rho)$ equals minus the sum of the negative eigenvalues of $\rho^{T_B}$. In the separable case $N(\rho) = 0$. For pure states the generalised concurrence~\cite{PhysRevA.64.042315} can also be used as entanglement measure. For a bipartite system $AB$, it is defined as
\begin{equation}
C^2 = 2 (1 - \operatorname{tr} \rho_A^2) \,,
\label{ec:conc}
\end{equation}
with $\rho_A$ the reduced density operator obtained by trace over the $B$ degrees of freedom. The result is the same when tracing over $\mathcal{H}_A$, and in the separable case $C^2 = 0$.

\subsection{Momentum-momentum entanglement}

We first address the entanglement between the two momenta. 
In order to numerically determine momentum-momentum entanglement we marginalise over spin degrees of freedom. Yet, the dimensionality of the space $\bar{\mathcal{H}}_{P_1} \otimes \bar{\mathcal{H}}_{P_2}$ is 64, 324 and 1024 for $\kappa = 2, 3, 4$, respectively. Numerical results for the entanglement measure, as given by the negativity $N$,  are presented in Table~\ref{tab:N1}. In this particular case, the numerical increase of $N$ with finer bin size does not mean `stronger entanglement' but only reflects the increased dimensionality of the spaces. 

\begin{table}[thb]
\begin{center}
\begin{tabular}{ccccccc}
& \multicolumn{3}{c}{$H \to WW$} & \multicolumn{3}{c}{$H \to ZZ$} \\ 
$\kappa$ & $2$ & $3$ & $4$ & $2$ & $3$ & $4$ \\
$N(P_1$-$P_2)$ & $3.29$ & $7.80$ & $14.1$ & $3.29$ & $7.81$ & $14.1$ \\
\end{tabular}
\caption{Entanglement  between the two momenta, computed after momentum discretisation with several binning choices.}
\label{tab:N1}
\end{center}
\end{table}

\subsection{Spin-momentum entanglement}

When addressing spin-momentum entanglement we consider the momentum space $\bar{\mathcal{H}}_{P} \equiv \bar{\mathcal{H}}_{P_1} \otimes \bar{\mathcal{H}}_{P_2}$ as a whole, and verify entanglement between the three possible bipartitions of $\bar{\mathcal{H}}_P \otimes \mathcal{H}_{S_1} \otimes \mathcal{H}_{S_2}$.
Furthermore, one can also investigate the entanglement between a pair of subsystems, when the third one is marginalised.
Tracing the full density operator ${\bar \rho}_{P S_1 S_2}$ over the Hilbert space of any of the subsystems $\bar{\mathcal{H}}_P$, $\mathcal{H}_{S_1}$, or $\mathcal{H}_{S_2}$, we obtain the reduced density operators for the other two, respectively ${\bar \rho}_{S_1 S_2}$, ${\bar \rho}_{P S_2}$, and ${\bar \rho}_{P S_1}$, and the entanglement between these subsystems can also be tested.

Intuitively, we expect that a finer momentum binning will result on larger spin-momentum entanglement, because the integration necessarily washes out some details of the momentum dependence. We investigate this effect in $H \to WW$ within a narrow mass slice $m_{V^*} \in [35,40]$ GeV, so that $\bar \rho_{P S_1 S_2}$ describes a pure state to an excellent approximation and the concurrence can be used as entanglement measure. (Results are alike for $H \to ZZ$.) We consider the three bipartitions $P$-$S_1 S_2$, $S_1$-$P S_2$ and $S_2$-$P S_1$, for which $C^2$ is computed from $\rho_{S_1 S_2}$, $\rho_{S_1}$ and $\rho_{S_2}$, respectively, using (\ref{ec:conc}). The maximum value of $C^2$ for these bipartitions obviously does not depend on the dimensionality of the discretised momentum space. Results are presented in Table~\ref{tab:C}, with the last column providing the maximum value of $C^2$ for that bipartition.

\begin{table}[thb]
\begin{center}
\begin{tabular}{ccccccc}
& \multicolumn{3}{c}{$H \to WW$}  \\ 
$\kappa$ & $2$ & $3$ & $4$ & max \\
$C^2(P$-$S_1 S_2)$ & $0.037$ & $0.060$ & $0.071$ & $16/9$ \\
$C^2(S_1$-$P S_2)$ & $4/3$ & $4/3$ & $4/3$ & $4/3$ \\
$C^2(S_2$-$P S_1)$ & $4/3$ & $4/3$ & $4/3$ & $4/3$ \\
\end{tabular}
\caption{Entanglement  between the two momenta in a slice $m_{V^*} \in [35,40]$ GeV, computed after momentum discretisation with several binning choices. The last column is the maximal value of the concurrence for the bipartition considered.}
\label{tab:C}
\end{center}
\end{table}

Besides the expected increase of $C^2(P$-$S_1 S_2)$ with $\kappa$, we observe that the entanglement is maximal between one spin and the rest of the system. The reason is that for isotropic Higgs decay, once the rest of degrees of freedom are integrated, the spin density operator is necessarily diagonal and maximally degenerate, with matrix elements equal to $1/3$.\footnote{With binned momenta this isotropy is broken if the two-dimensional bins span different solid angles.} This density operator has the lowest value of $\operatorname{tr} \rho_A^2$ in the concurrence definition (\ref{ec:conc}) and thus maximal entanglement. In all these cases the negativity is $N = 1$.

In full decay phase space, the produced state is not pure and we use $N$ to measure entanglement between different pairs of subsystems of the full $\bar{\mathcal{H}}_P \otimes \mathcal{H}_{S_1} \otimes \mathcal{H}_{S_2}$ space.
Numerical results  are presented in Table~\ref{tab:N}, using several binning choices.

\begin{table}[thb]
\begin{center}
\begin{tabular}{ccccccc}
& \multicolumn{3}{c}{$H \to WW$} & \multicolumn{3}{c}{$H \to ZZ$} \\ 
$\kappa$ & $2$ & $3$ & $4$ & $2$ & $3$ & $4$ \\
$N(P$-$S_1 S_2)$ & $0.348$ & $0.617$ & $0.671$ & $0.353$ & $0.625$ & $0.680$ \\
$N(S_1$-$P S_2)$ & $0.998$ & $0.998$ & $0.998$ & $0.998$ & $0.998$ & $0.998$ \\
$N(S_2$-$P S_1)$ & $0.998$ & $0.998$ & $0.998$ & $0.998$ & $0.998$ & $0.998$ \\
$N(S_1$-$S_2)$     & $0.923$ & $0.884$ & $0.868$ & $0.923$ & $0.884$ & $0.868$ \\
$N(P$-$S_1)$           & $0.022$ & $0.075$ & $0.086$ & $0.023$ & $0.076$ & $0.088$ \\
$N(P$-$S_2)$           & $0.022$ & $0.075$ & $0.086$ & $0.023$ & $0.076$ & $0.088$
\end{tabular}
\caption{Entanglement  between different subsystems, computed after momentum discretisation with several binning choices.}
\label{tab:N}
\end{center}
\end{table}

The entanglement measures $N(P$-$S_1 S_2)$, $N(P$-$S_1)$ and $N(P$-$S_2)$ increase with a finer binning as expected, although the effect cannot be fully atributed to `larger entanglement' because of the different dimensionalities of momentum space.  
The entanglement measures $N(S_1$-$P S_2)$ and $N(S_2$-$P S_1)$ are quite close to unity in all cases.  The reason is that the helicity amplitudes $\app$, $\azz$, $\amm$ have a mild dependence on $m_{V^*}$; therefore, the density operator $\rho_{P S_1 S_2}$ describes a nearly-pure state in which case one has $N=1$ as previously seen. 

The spin entanglement measure $N(S_1$-$S_2)$ can be computed by integrating $\rho_{PS_1 S_2}$ over momenta, without the need of discretisation, and has also been obtained in Ref.~\cite{Aguilar-Saavedra:2024whi} by tracing the density operator $\rho_{L S_1 S_2}$ over o.a.m.\ degrees of freedom. Its value is $N(S_1$-$S_2) = 0.843$ for both $H \to WW$ and $H \to ZZ$.
The comparison with the values in Table~\ref{tab:N}, obtained with momentum discretisation, suggests that $\kappa = 4$ provides a sufficiently fine binning.

Overall, from the results in Table~\ref{tab:N} we observe that the state is fully inseparable with respect to the three bipartitions, since $N(P$-$S_1 S_2)$, $N(S_1$-$P S_2)$ and $N(S_2$-$P S_1)$ are all nonzero. Moreover, for a single weak boson its momentum and spin are entangled,  $N(P$-$S_1) = N(P$-$S_2) \neq 0$.

\section{Experimental prospects for $H \to ZZ$}
\label{sec:5}

 In this section we assess the statistical sensitivity with which entanglement measures can be inferred in $pp \to H \to ZZ \to 4 \ell$, under the assumption that the $H \to ZZ$ decay is described by the standard coherent QFT state (\ref{ec:QFTdec}).
The $H \to ZZ \to 4 \ell$ decay mode is very clean, though with a small branching ratio. Experimental uncertainties are dominated by the statistical ones. We assess in this section the statistical uncertainty in the determination of various entanglement measures in $pp \to H \to ZZ \to 4 \ell$  the LHC, using Run 2$+$3 data, and at the HL-LHC. 

We do not include backgrounds in our analysis. The leading one is the electroweak process $pp \to ZZ/Z\gamma \to 4\ell$, which is about 4 times smaller at the Higgs peak~\cite{ATLAS:2023mqy,CMS:2021ugl}. Although a background subtraction is necessary to obtain the relevant signal distributions, the main effect of the presence of this small background is a slight increase in the statistical uncertainty of the measurement. Also, next-to-leading order corrections to the Higgs decay can be subtracted with minimal impact on the statistical uncertainty~\cite{Aguilar-Saavedra:2025byk,Aguilar-Saavedra:2026wuq}.

In the same-flavour channels $H \to ZZ \to 4e/4\mu$, an additional complication arises from identical-particle effects~\cite{Aguilar-Saavedra:2024jkj} which prevent a description in terms of two intermediate spin-1 bosons. A lower cut on the highest opposite-sign lepton invariant mass removes the interference between Feynman diagrams~\cite{Aguilar-Saavedra:2024whi} thereby allowing the system to be effectively treated as a pair of spin-1 particles. The efficiency of that cut, approximately 0.7, is taken into account in our sensitivity estimates. 

For the calculation of the expected number of events we use state-of-the art values of the Higgs production cross section and branching ratio into four electrons or muons. The cross section at next-to-next-to-next-to-leading order is 48.61 pb, 52.23 pb and 54.67 pb at centre-of-mass energies of 13, 13.6 and 14 TeV~\cite{Cepeda:2019klc}, and the Higgs branching ratios into $ee\mu\mu$ and $4e/4\mu$ are $5.9 \times 10^{-5}$ and $6.5 \times 10^{-5}$, respectively~\cite{LHCHiggsCrossSectionWorkingGroup:2016ypw}.
The assumed luminosities are 350 fb$^{-1}$ for Runs 2+3 and 3 ab$^{-1}$ for HL-LHC. In order to have a more realistic estimate of the number of events in each case a lepton detection efficiency of 0.7 is assumed, yielding an overall detection efficiency of 0.25. This efficiency accounts for the minimum transverse momentum ($p_T$) thresholds required for lepton detection. We do not include any trigger requirement. The presence of four leptons from the Higgs decay, some of them with significant $p_T$, is expected to fulfill one or many of the trigger conditions for one, two, or three leptons~\cite{trigger}. In addition, we include the efficiency of the invariant mass cut required to remove the interference in same-flavour final states. Overall, the expected number of events for Runs 2+3 and HL-LHC are $N = 490$ and $N = 4500$, respectively.

The statistical uncertainty is estimated by performing pseudo-experiments. In each pseudo-experiment, a subset of $N$ random events is drawn from the total event set (7 millions), and for this subset the discretised density operator  is calculated as discussed in section~\ref{sec:3}. Because the number of events in the samples is not large even for HL-LHC, we use three $m_{V^*}$ bins of 20 GeV, which provides an approximation that is accurate enough. Once the density operator is obtained,  the entanglement measures $N$ for different subsystems are obtained as outlined in section~\ref{sec:4}. A large number of $2 \times 10^4$ pseudo-experiments is performed in order to obtain the probability density function (p.d.f.) of these quantities.

We have investigated the sensitivity for the three binning options $\kappa = 2$, $3$ and $4$. For momentum-momentum entanglement the sensitivity is similar in all cases: even if the entanglement measure increases as seen in Table~\ref{tab:N1},  the statistical uncertainties also increase by the same amount. This suggests that the variation of $N$ with $\kappa$ is a mere scaling due to the increasing dimensionality. On the other hand, the sensitivity to spin-momentum entanglement increases with $\kappa$. This is also understood from the results for the concurrence in Table~\ref{tab:C}, which unambiguously indicate larger entanglement for finer bin sizes. For brevity, we only present results using $\kappa = 2$ and $\kappa = 4$ for momentum-momentum and spin-momentum entanglement, respectively.

\begin{figure}[htb]
\begin{center}
\begin{tabular}{c}
\includegraphics[height=5.5cm,clip=]{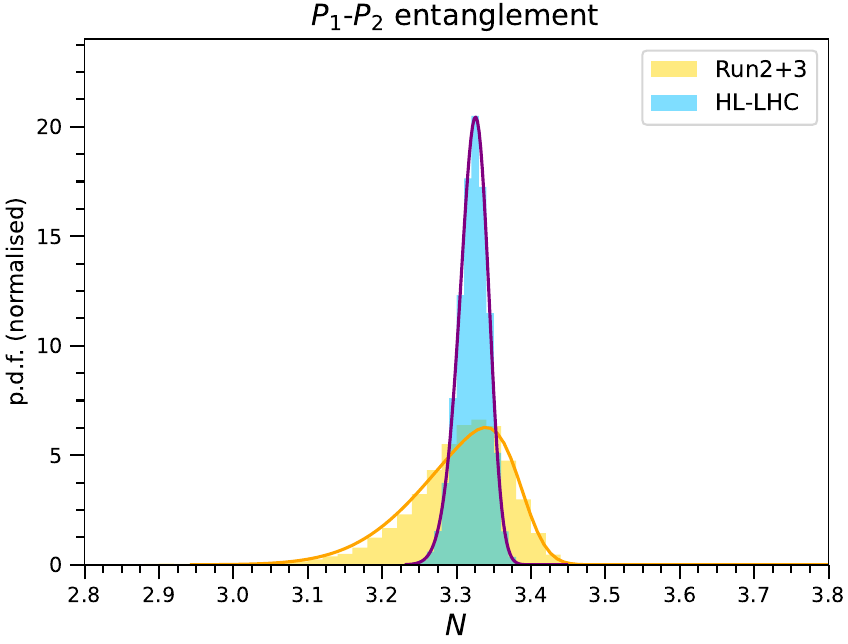} \\
\includegraphics[height=5.5cm,clip=]{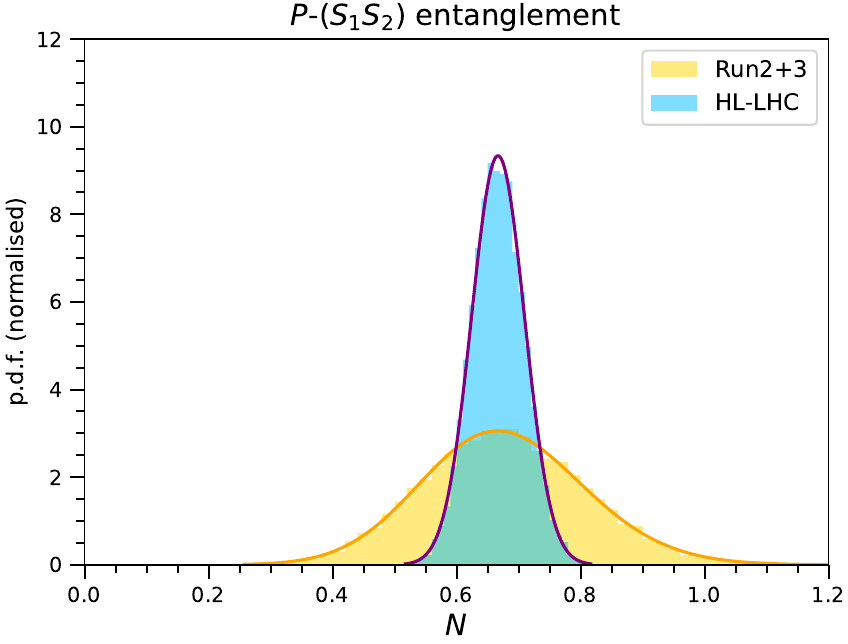} \\
\includegraphics[height=5.5cm,clip=]{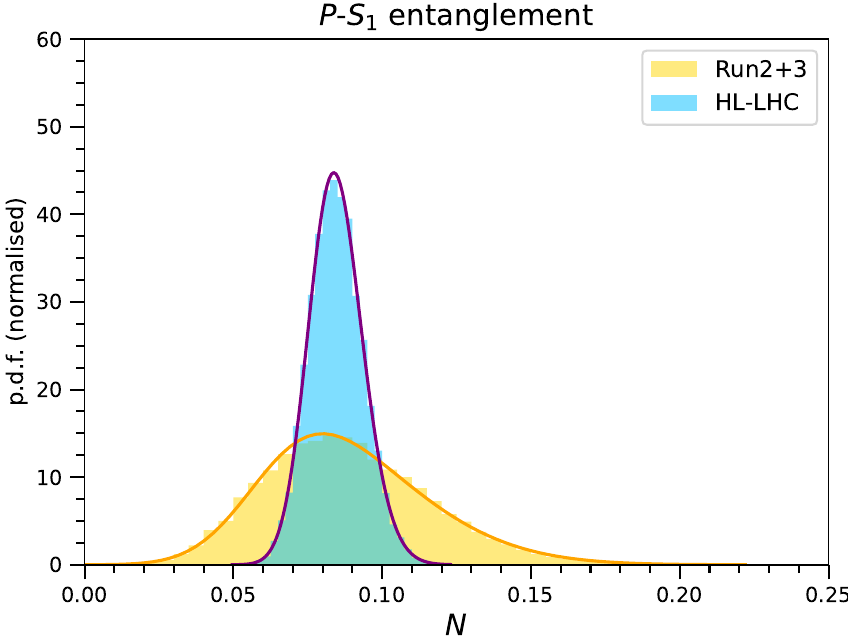}
\end{tabular}
\caption{Probability density functions for momentum entanglement measures, as obtained from the pseudo-experiments. The solid lines represent the best-fit Gaussian or skew-normal distributions.}
\label{fig:pdf}
\end{center}
\end{figure}

The central values and statistical uncertainties obtained from the pseudo-experiments are reported in Table~\ref{tab:Nexp}. The SM theoretical prediction and value of $\kappa$ used are provided in the last two columns. The cases where the p.d.f. deviates appreciably from a Gaussian are marked with an asterisk. 
\begin{table}[tb]
\begin{center}
\begin{tabular}{cccccccc}
 & Run 2$+$3 & HL-LHC & SM value & $\kappa$ \\
$N(P_1$-$P_2)$ & $3.30 \pm 0.07^*$ & $3.32 \pm 0.02$ & 3.29 & 2 \\
$N(P$-$S_1 S_2)$ & $0.68 \pm 0.13$ & $0.67 \pm 0.04$ & $0.680$ & 4 \\
$N(S_1$-$P S_2)$ & $0.997 \pm 0.003$ & $0.998 \pm 0.001$ & $0.998$ & 4 \\
$N(S_2$-$P S_1)$ & $0.997 \pm 0.003$ & $0.998 \pm 0.001$ & $0.998$ & 4 \\
$N(S_1$-$S_2)$     & $0.87 \pm 0.04^*$ & $0.881 \pm 0.012$ & $0.868$ & 4 \\
$N(P$-$S_1)$           & $0.088 \pm 0.027^*$ & $0.085 \pm 0.009$ & $0.088$ & 4 \\
$N(P$-$S_2)$           & $0.088 \pm 0.027^*$ & $0.085 \pm 0.009$ & $0.088$ & 4
\end{tabular}
\caption{Expected statistical uncertainty for entanglement measures between different subsystems.}
\label{tab:Nexp}
\end{center}
\end{table}
The p.d.f. of $N(P_1$-$P_2)$, $N(P$-$S_1 S_2)$ and $N(P$-$S_1)$ are shown in Fig.~\ref{fig:pdf}. As discussed in section~\ref{sec:4}, 
for this decay the entanglement between one spin and the rest of the system is nearly maximal, which in the pseudo-experiments manifests as only tiny deviations from unity. The spin-spin entanglement $N(S_1$-$S_2)$ is also of high interest but it can already be computed without momentum discretisation.

Our estimates indicate that, using Run 2$+$3 data, the inferred entanglement measures for the two momenta and for momentum and the two spins exceed 5$\sigma$.  After marginalising over the spin degrees of freedom of one $Z$ boson, the significance for momentum-spin entanglement of the other boson exceeds 3$\sigma$.  At the HL-LHC all the entanglement measures considered reach significances above $5\sigma$.

We also note that the central values obtained from the pseudo-experiments show very good agreement with the SM prediction in all cases, indicating that the use of three $m_{V^*}$ bins of 20 GeV is appropriate and does not introduce any significant bias. Employing smaller bin sizes increases the statistical uncertainties and ultimately compromises the accuracy of the parameter determination.

\section{Final remarks}

Particle detection at colliders projects onto momentum eigenstates, thereby preventing a direct measurement of the interference between amplitudes with different momenta. However, in the standard coherent $S$-matrix description of the decay, these interference terms are fixed by the decay amplitudes. Here we have parametrised the amplitudes for $H \to WW,ZZ$ in terms of three quantities $\app$, $\azz$, $\amm$, which can be determined from experimentally accessible observables. This construction relies only on quantum mechanics, angular momentum conservation and the assumptions of spin zero for the Higgs boson and spin one for $W$ and $Z$. Otherwise, it does not assume any specific structure for the interactions mediating the decay.

A determination of $\app$, $\azz$, $\amm$ from data only relying on the above general assumptions is, in principle, feasible~\cite{Aguilar-Saavedra:2024whi}. However, with current statistics it leads to sizeable uncertainties. Here we have estimated the expected statistical precision for entanglement measures in $H \to ZZ$ assuming CP conservation in the decay as well. Under this mild assumption, supported by current measurements and in agreement with the SM expectation, the LHC has excellent sensitivity to momentum entanglement measures. In particular, with the existing dataset the momentum-spin entanglement measure of a single $Z$ boson can be determined with a statistical significance above $3\sigma$, while the measures associated with momentum-momentum and momentum-spin entanglement exceed $5\sigma$. At the HL-LHC all the entanglement measures considered can be determined with a statistical significance exceeding $5\sigma$.

 It is important to stress the role of the coherent QFT description in this inference. As mentioned in the introduction, particle detection prevents the direct measurement of interference terms involving different momenta. The observables considered here determine the amplitudes entering the construction of the density operator, but do not independently probe coherence between distinct momentum configurations. A completely momentum-decohered state analogue to (\ref{ec:rhoslice}) retaining the same diagonal momentum distributions and momentum-conditioned spin density matrices would not be distinguished by the observables considered here. Therefore, the quoted sensitivities refer to the entanglement measures of the coherent QFT state, rather than to a model-independent exclusion of all momentum-separable density operators.

The discretisation method introduced here is broadly applicable to collider processes where momentum becomes entangled with other degrees of freedom. The only process-specific element is whether the amplitudes can be expressed in terms of measurable observables. When this condition is met, collider experiments gain access to a new class of quantum correlations, opening the door to systematic studies of spin-momentum entanglement at the energy frontier.

\section*{Acknowledgements}

I thank J.A. Casas for useful comments.
This work has been supported by the Spanish Research Agency (Agencia Estatal de Investigaci\'on) through projects PID2022-142545NB-C21 and CEX2020-001007-S funded by MCIN/AEI/10.13039/501100011033.

\appendix
\section{Determination of $\app$, $\azz$, $\amm$ }
\label{sec:a}

The determination of the helicity amplitudes $\app$, $\azz$ and $\amm$ from (pseudo-)data has been described in detail previously~\cite{Aguilar-Saavedra:2024whi}, and we refer the reader to that reference for a comprehensive discussion. An extraction of these parameters, including their relative phases, is possible relying only on quantum mechanics, angular momentum conservation and the assumptions of spin zero for the Higgs boson and spin one for $W$ and $Z$. Here we make the additional assumption of CP conservation, which makes the determination from data is simpler. 

For the decay $H \to V_1 V_2 \to f_1 f_1' f_2 f_2'$ we denote as $(\theta_{1,2}^*,\varphi_{1,2}^*)$ the polar and azimuthal angles of $f_{1,2}$ in the $V_{1,2}$ rest frame, defined with respect to a coordinate system where the $\hat z$ axis is taken along the direction of the $V_1$ three-momentum in the Higgs rest frame. (The precise orientation of the two other axes is unimportant for our purposes.) The four-dimensional angular distribution of the two fermions $f_{1,2}$ can be expanded in spherical harmonics~\cite{Aguilar-Saavedra:2022wam,Aguilar-Saavedra:2022mpg}
\begin{align}
\frac{1}{\sigma}\frac{d\sigma}{d\Omega_1^* d\Omega_2^*} & = \frac{1}{(4\pi)^2}\left[ 1 + \mathcal{A}_{lm}^1 Y_l^m(\theta_1^*, \varphi_1^*) \right. \notag \\
& +\mathcal{A}_{lm}^2 Y_l^m(\theta_2^*, \varphi_2^*) . \notag \\
& \left. + \mathcal{C}_{l_1 m_1 l_2 m_2} Y_{l_1}^{m_1}(\theta_1^*, \varphi_1^*)Y_{l_2}^{m_2}(\theta_2^*, \varphi_2^*)  \right] \,,
\label{ec:dist4D}
\end{align}
with $\mathcal{A}_{lm}^{1,2}$, $\mathcal{C}_{l_1 m_1 l_2 m_2}$ independent of the angles. Using the helicity formalism, one finds~\cite{Aguilar-Saavedra:2022mpg}
\begin{equation}
\mathcal{A}_{20}^1 = \mathcal{A}_{20}^2 = \sqrt{\frac{\pi}{5}} \frac{1}{\mathcal{N}} \left[ 
|a_{1 1}|^2 + |a_{\smo \smo}|^2 - 2 |a_{00}|^2 \right] \,,
\end{equation}
with $\mathcal{N} = |\app|^2 + |\azz|^2 + |\amm|^2$.\footnote{Since the angular distributions and density operator are independent of this normalisation, we can set $\mathcal{N} = 1$ for simplicity.} 
The parameters $\mathcal{A}_{20}^1$ and $\mathcal{A}_{20}^2$ can be measured in the one-dimensional distributions of $\theta_1^*$ and $\theta_2^*$, respectively, since
\begin{equation}
Y_2^0 (\theta,\varphi) = \frac{1}{4} \sqrt{\frac{5}{\pi}} (3 \cos^2 \theta - 1 ) \,,
\end{equation}
which is independent of the azimuthal angles.
In the CP-conserving case $\app = \amm$, therefore the moduli of the three amplitudes can be determined from either $\mathcal{A}_{20}^1$ or $\mathcal{A}_{20}^2$, and the two statistically-independent measurements can be combined. The relative sign between $\app$ and $\azz$ is fixed by the Lorentz structure of the $H V_1 V_2$ vertex~\cite{Aguilar-Saavedra:2022wam}.


\begin{thebibliography}{99}

\bibitem{Schrodinger:1935}
E. Schrödinger,
Mathematical Proceedings of the Cambridge Philosophical Society. {\bf 31} (4) (1935), 555–563.


\bibitem{Afik:2020onf}
Y.~Afik and J.~R.~M.~de Nova,
Eur. Phys. J. Plus \textbf{136} (2021) no.9, 907
[arXiv:2003.02280 [quant-ph]].

\bibitem{Severi:2021cnj}
C.~Severi, C.~D.~Boschi, F.~Maltoni and M.~Sioli,
Eur. Phys. J. C \textbf{82} (2022) no.4, 285
[arXiv:2110.10112 [hep-ph]].

\bibitem{Afik:2022kwm}
Y.~Afik and J.~R.~M.~de Nova,
Quantum \textbf{6} (2022), 820
[arXiv:2203.05582 [quant-ph]].

\bibitem{Aoude:2022imd}
R.~Aoude, E.~Madge, F.~Maltoni and L.~Mantani,
Phys. Rev. D \textbf{106} (2022) no.5, 055007
[arXiv:2203.05619 [hep-ph]].

\bibitem{Aguilar-Saavedra:2022uye}
J.~A.~Aguilar-Saavedra and J.~A.~Casas,
Eur. Phys. J. C \textbf{82} (2022) no.8, 666
[arXiv:2205.00542 [hep-ph]].

\bibitem{Severi:2022qjy}
C.~Severi and E.~Vryonidou,
JHEP \textbf{01} (2023), 148
[arXiv:2210.09330 [hep-ph]].

\bibitem{Dong:2023xiw}
Z.~Dong, D.~Gon\c{c}alves, K.~Kong and A.~Navarro,
Phys. Rev. D \textbf{109} (2024) no.11, 115023
[arXiv:2305.07075 [hep-ph]].

\bibitem{Han:2023fci}
T.~Han, M.~Low and T.~A.~Wu,
JHEP \textbf{07} (2024), 192
[arXiv:2310.17696 [hep-ph]].

\bibitem{Maltoni:2024tul}
F.~Maltoni, C.~Severi, S.~Tentori and E.~Vryonidou,
JHEP \textbf{03} (2024), 099
[arXiv:2401.08751 [hep-ph]].

\bibitem{Maltoni:2024csn}
F.~Maltoni, C.~Severi, S.~Tentori and E.~Vryonidou,
JHEP \textbf{09} (2024), 001
[arXiv:2404.08049 [hep-ph]].

\bibitem{Cheng:2024btk}
K.~Cheng, T.~Han and M.~Low,
Phys. Rev. D \textbf{111} (2025) no.3, 033004
[arXiv:2407.01672 [hep-ph]].

\bibitem{Aoude:2025ovu}
R.~Aoude, A.~J.~Barr, F.~Maltoni and L.~Satrioni,
[arXiv:2504.07030 [quant-ph]].



\bibitem{Aguilar-Saavedra:2022wam}
J.~A.~Aguilar-Saavedra, A.~Bernal, J.~A.~Casas and J.~M.~Moreno,
Phys. Rev. D \textbf{107} (2023) no.1, 016012
[arXiv:2209.13441 [hep-ph]].

\bibitem{Ashby-Pickering:2022umy}
R.~Ashby-Pickering, A.~J.~Barr and A.~Wierzchucka,
JHEP \textbf{05} (2023), 020
[arXiv:2209.13990 [quant-ph]].

\bibitem{Aguilar-Saavedra:2022mpg}
J.~A.~Aguilar-Saavedra,
Phys. Rev. D \textbf{107} (2023) no.7, 076016
[arXiv:2209.14033 [hep-ph]].

\bibitem{Fabbrichesi:2023cev}
M.~Fabbrichesi, R.~Floreanini, E.~Gabrielli and L.~Marzola,
Eur. Phys. J. C \textbf{83} (2023) no.9, 823
[arXiv:2302.00683 [hep-ph]].

\bibitem{Morales:2023gow}
R.~A.~Morales,
Eur. Phys. J. Plus \textbf{138} (2023) no.12, 1157
[arXiv:2306.17247 [hep-ph]].

\bibitem{Aoude:2023hxv}
R.~Aoude, E.~Madge, F.~Maltoni and L.~Mantani,
JHEP \textbf{12} (2023), 017
[arXiv:2307.09675 [hep-ph]].

\bibitem{Bernal:2023ruk}
A.~Bernal, P.~Caban and J.~Rembieli{\'n}ski,
Eur. Phys. J. C \textbf{83} (2023) no.11, 1050
[arXiv:2307.13496 [hep-ph]].

\bibitem{Bernal:2024xhm}
A.~Bernal, P.~Caban and J.~Rembieli{\'n}ski,
Sci. Rep. \textbf{15} (2025) no.1, 23410
[arXiv:2405.16525 [hep-ph]].

\bibitem{Ruzi:2024cbt}
A.~Ruzi, Y.~Wu, R.~Ding, S.~Qian, A.~M.~Levin and Q.~Li,
JHEP \textbf{10} (2024), 211
[arXiv:2408.05429 [hep-ph]].

\bibitem{Grossi:2024jae}
M.~Grossi, G.~Pelliccioli and A.~Vicini,
JHEP \textbf{12} (2024), 120
[arXiv:2409.16731 [hep-ph]].

\bibitem{Wu:2024ovc}
Y.~Wu, R.~Jiang, A.~Ruzi, Y.~Ban, X.~Yan and Q.~Li,
Phys. Rev. D \textbf{111} (2025) no.3, 036008
[arXiv:2410.17025 [hep-ph]].

\bibitem{Bernal:2025zqq}
A.~Bernal, J.~A.~Casas and J.~Falceto,
Phys. Rev. D \textbf{112} (2025) no.1, 016024
[arXiv:2503.17297 [quant-ph]].

\bibitem{DelGratta:2025qyp}
M.~Del Gratta, F.~Fabbri, P.~Lamba, F.~Maltoni and D.~Pagani,
JHEP \textbf{09} (2025), 013
[arXiv:2504.03841 [hep-ph]].

\bibitem{Ding:2025mzj}
R.~Ding, A.~Ruzi, S.~Qian, A.~Levin, Y.~Wu and Q.~Li,
[arXiv:2504.09832 [hep-ph]].

\bibitem{Goncalves:2025qem}
D.~Gon{\c{c}}alves, A.~Kaladharan, F.~Krauss and A.~Navarro,
[arXiv:2505.12125 [hep-ph]].

\bibitem{Goncalves:2025xer}
D.~Gon{\c{c}}alves, A.~Kaladharan and A.~Navarro,
[arXiv:2506.19951 [hep-ph]].




\bibitem{Altakach:2022ywa}
M.~M.~Altakach, P.~Lamba, F.~Maltoni, K.~Mawatari and K.~Sakurai,
Phys. Rev. D \textbf{107} (2023) no.9, 093002
[arXiv:2211.10513 [hep-ph]].


\bibitem{Ehataht:2023zzt}
K.~Ehat\"aht, M.~Fabbrichesi, L.~Marzola and C.~Veelken,
Phys. Rev. D \textbf{109} (2024) no.3, 032005
[arXiv:2311.17555 [hep-ph]].

\bibitem{Fabbrichesi:2024wcd}
M.~Fabbrichesi and L.~Marzola,
Phys. Rev. D \textbf{110} (2024) no.7, 076004
[arXiv:2405.09201 [hep-ph]].

\bibitem{Han:2025ewp}
T.~Han, M.~Low and Y.~Su,
JHEP \textbf{10} (2025), 217
[arXiv:2501.04801 [hep-ph]].

\bibitem{Zhang:2025mmm}
Y.~Zhang, B.~H.~Zhou, Q.~B.~Liu, S.~Li, S.~C.~Hsu, T.~Han, M.~Low and T.~A.~Wu,
[arXiv:2504.01496 [hep-ph]].



\bibitem{Afik:2025grr}
Y.~Afik, Y.~Kats, J.~R.~M.~de Nova, A.~Soffer and D.~Uzan,
Phys. Rev. D \textbf{111} (2025) no.11, L111902
[arXiv:2406.04402 [hep-ph]].






\bibitem{Aguilar-Saavedra:2023hss}
J.~A.~Aguilar-Saavedra,
Phys. Rev. D \textbf{108} (2023) no.7, 076025
[arXiv:2307.06991 [hep-ph]].

\bibitem{Aguilar-Saavedra:2024fig}
J.~A.~Aguilar-Saavedra and J.~A.~Casas,
Phys. Rev. Lett. \textbf{133} (2024) no.11, 111801
[arXiv:2401.06854 [hep-ph]].

\bibitem{Aguilar-Saavedra:2024hwd}
J.~A.~Aguilar-Saavedra,
Phys. Rev. D \textbf{109} (2024) no.9, 096027
[arXiv:2401.10988 [hep-ph]].




\bibitem{Aguilar-Saavedra:2024vpd}
J.~A.~Aguilar-Saavedra,
Phys. Lett. B \textbf{855} (2024), 138849
[arXiv:2402.14725 [hep-ph]].

\bibitem{Aguilar-Saavedra:2024whi}
J.~A.~Aguilar-Saavedra,
Phys. Rev. D \textbf{109} (2024) no.11, 113004
[arXiv:2403.13942 [hep-ph]].


\bibitem{Peres:1996dw}
A.~Peres,
Phys. Rev. Lett. \textbf{77} (1996), 1413-1415
[arXiv:quant-ph/9604005 [quant-ph]].

\bibitem{Horodecki:1997vt}
P.~Horodecki,
Phys. Lett. A \textbf{232} (1997), 333
[arXiv:quant-ph/9703004 [quant-ph]].

\bibitem{Jacob:1959at}
M.~Jacob and G.~C.~Wick,
Annals Phys. \textbf{7}, 404-428 (1959)

\bibitem{Berge:2015jra}
S.~Berge, S.~Groote, J.~G.~K\"orner and L.~Kaldam\"ae,
Phys. Rev. D \textbf{92}, no.3, 033001 (2015)
[arXiv:1505.06568 [hep-ph]].

\bibitem{Alwall:2014hca}
J.~Alwall, R.~Frederix, S.~Frixione, V.~Hirschi, F.~Maltoni, O.~Mattelaer, H.~S.~Shao, T.~Stelzer, P.~Torrielli and M.~Zaro,
JHEP \textbf{07}, 079 (2014)
[arXiv:1405.0301 [hep-ph]].


\bibitem{Plenio:2007zz}
M.~B.~Plenio and S.~Virmani,
Quant. Inf. Comput. \textbf{7}, no.1-2, 001-051 (2007)
[arXiv:quant-ph/0504163 [quant-ph]].

\bibitem{PhysRevA.64.042315}
P. Rungta, V. Bu\v{z}ek, C.M. Caves, M. Hillery and G. J. Milburn,
Phys. Rev. A \textbf{64} no. 4, 042315 (2001)
[arXiv:quant-ph/0102040].

\bibitem{ATLAS:2023mqy}
G.~Aad \textit{et al.} [ATLAS],
JHEP \textbf{05} (2024), 105
[arXiv:2304.09612 [hep-ex]].

\bibitem{CMS:2021ugl}
A.~M.~Sirunyan \textit{et al.} [CMS],
Eur. Phys. J. C \textbf{81} (2021) no.6, 488
[arXiv:2103.04956 [hep-ex]].




\bibitem{Aguilar-Saavedra:2025byk}
J.~A.~Aguilar-Saavedra,
Eur. Phys. J. C \textbf{85} (2025) no.9, 969
[arXiv:2505.11870 [hep-ph]].

\bibitem{Aguilar-Saavedra:2026wuq}
J.~A.~Aguilar-Saavedra and P.~P.~Giardino,
[arXiv:2603.11288 [hep-ph]].


\bibitem{Aguilar-Saavedra:2024jkj}
J.~A.~Aguilar-Saavedra,
Phys. Lett. B \textbf{868} (2025), 139639
[arXiv:2411.13464 [hep-ph]].

\bibitem{Cepeda:2019klc}
M.~Cepeda, S.~Gori, P.~Ilten, M.~Kado, F.~Riva, R.~Abdul Khalek, A.~Aboubrahim, J.~Alimena, S.~Alioli and A.~Alves, \textit{et al.}
CERN Yellow Rep. Monogr. \textbf{7}, 221-584 (2019)
[arXiv:1902.00134 [hep-ph]].

\bibitem{LHCHiggsCrossSectionWorkingGroup:2016ypw}
D.~de Florian \textit{et al.} [LHC Higgs Cross Section Working Group],
CERN Yellow Rep. Monogr. \textbf{2} (2017), 1-869
[arXiv:1610.07922 [hep-ph]].

\bibitem{trigger}
ATLAS Collaboration, Trigger menu in 2017, Technical Report No. ATL-DAQ-PUB-2018-002, CERN, Geneva, 2018.










\end{thebibliography}
\end{document}